\documentclass[12pt]{article}
\usepackage{epsfig}

\def\etal{{\it et al.}}

\def\sone{\pi^- 6\pi^0}
\def\sthree{2\pi^- \pi^+ 4\pi^0}
\def\sfive{3\pi^- 2\pi^+ 2\pi^0}
\def\sseven{4\pi^- 3\pi^+}

\def\taufivea{\tau^- \rightarrow  3\pi^- 2\pi^+ \nu_{\tau} } 
\def\taufiveb{\tau^- \rightarrow  2\pi^- \pi^+ 2\pi^0 \nu_{\tau} } 
\def\taufivec{\tau^- \rightarrow  \pi^-  4\pi^0 \nu_{\tau} } 

\def\sixa{\tau^- \rightarrow  3\pi^- 2\pi^+ \pi^0 \nu_{\tau} } 
\def\sixb{\tau^- \rightarrow  2\pi^- \pi^+ 3\pi^0 \nu_{\tau} } 
\def\sixc{\tau^- \rightarrow  \pi^-  5\pi^0 \nu_{\tau} } 

\def\sevena{\tau^- \rightarrow  4\pi^- 3\pi^+  \nu_{\tau} } 
\def\sevenb{\tau^- \rightarrow  3\pi^- 2\pi^+ 2\pi^0 \nu_{\tau} } 
\def\sevenc{\tau^- \rightarrow  2\pi^- \pi^+ 4\pi^0 \nu_{\tau} } 
\def\sevend{\tau^- \rightarrow  \pi^-  6\pi^0 \nu_{\tau} } 

\def\omga{\tau^- \rightarrow  \pi^- \omega \pi^0 \nu_{\tau} } 
\def\omgb{\tau^- \rightarrow  \pi^- \omega 2\pi^0 \nu_{\tau} } 
\def\omgc{\tau^- \rightarrow  2\pi^- \pi^+ \omega  \nu_{\tau} } 

\begin{document}
\begin{flushright}
UVIC-HEP-98-01 \\
October 19, 1998
\end{flushright}

\begin{center}
{\bf\large The Isospin Model prediction for multi-pion tau decays}  \\
\vspace{1cm}
{Randall J. Sobie}  \\
{The Institute of Particle Physics of Canada and 
 The University of Victoria \\
 Department of Physics and Astronomy, \\
 P.O. Box 3055, Victoria, British Columbia, \\
 Canada, V8W 3P6} \\
\end{center}
\vspace{2cm}
\begin{center}
Abstract
\end{center}
The predictions of an isospin model are compared with the 
branching ratios of the 5 and 6 pion decays of the tau lepton.
In both cases, the isospin model suggests that the tau 
favours decays in which there is an $\omega$ resonance.
Recent measurements of such tau decays confirm this hypothesis.
If the decay of the tau to 7 pions also proceeds through an intermediate 
$\omega$, then the isospin model predicts that the branching
ratio of the $\sevena$ should be small when compared with
other 7 pion decays.
New limits on this mode appear to support this argument.

\newpage

The large samples of tau leptons at CLEO and LEP have made it
possible to measure the branching ratio of rare multi-pion modes.
Further, advances in the understanding of the detectors have helped
to identify the underlying resonances, such as the $\eta$ and 
$\omega$, in these decay modes \cite{heltsley}.
The isospin model  \cite{pais} can been used to predict
the relative branching ratios of the 5 and 6 pion
decay modes  \cite{sobie, rouge}.
We re-examine the predictions in light of the new branching
ratio measurements and find that the isospin model predicts that
decays involving $\omega$ resonances should be favoured.
If the decay of the tau to 7 pions also proceeds through an intermediate 
$\omega$, then the isospin model may explain
the non-observation of the $\sevena$ decay.

The isospin model can describe the decay of the tau
into multi-pion final states.
A system with $N$ pions can be expressed in terms of orthogonal states
labeled by three integers $(N_1~N_2~N_3)$,
where $N = N_1 + N_2 + N_3$ and $N_1 \geq N_2 \geq N_3$.
The numbers $(N_1~N_2~N_3)$ give the internal pion symmetries:
$N_3$ is the number of $3\pi$ states with $I=0$ ($\omega$) states;
$N_2-N_3$ gives the number of $2\pi$ states with $I=1$ ($\rho$) states;
and $N_1-N_2$ gives the number of $\pi$ states.
The different states can be associated with the different
dynamical mechanisms (see table~\ref{table:modes}).
The isospin model does not predict the relative strength of
each state for a given number of pions.
Consequently only limits on the relative branching ratios can be predicted.
The relative branching ratio is defined to be the ratio of the
width of a particular $N$-pion tau decay channel to the total width
of all the $N$-pion tau decay channels.
The coefficients relating the isospin states to the observed final
states in 5 and 6 pion tau decays can be found in \cite{sobie}.

\begin{figure}
\begin{center}
\mbox{\epsfig{file=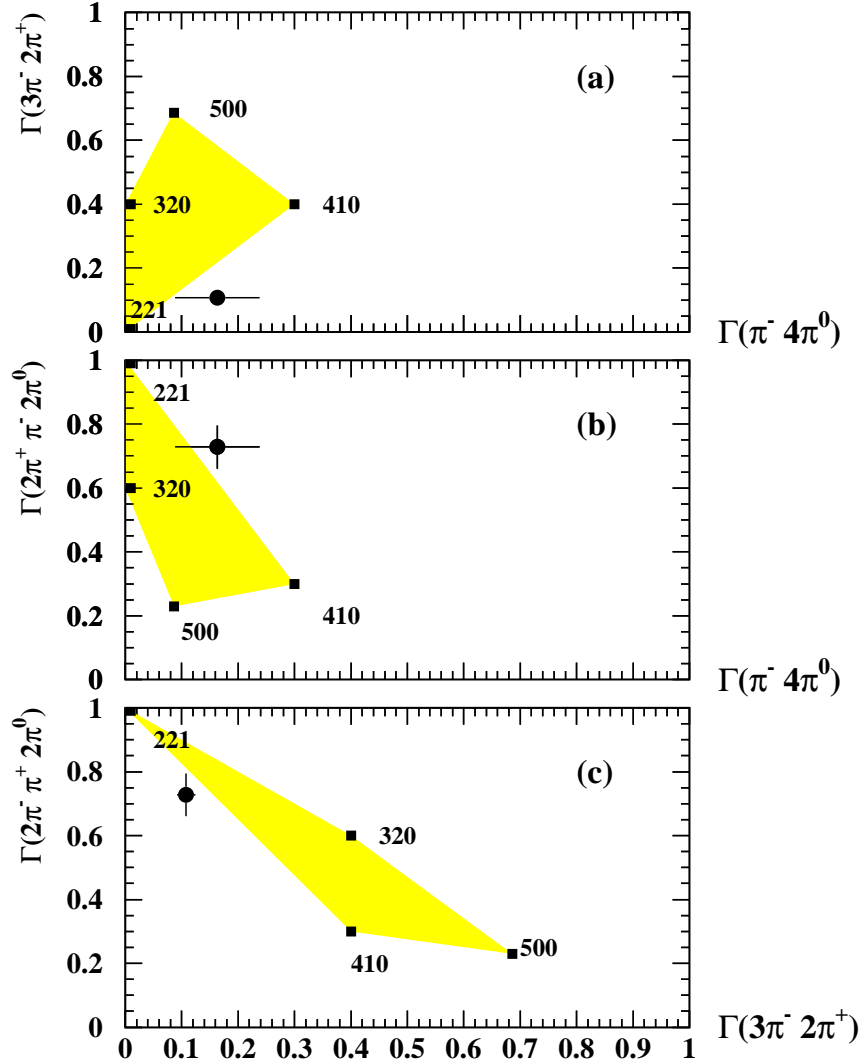,height=15cm}}
\end{center}
\caption[]{The isospin  model prediction is plotted for the 5 pion 
$\tau$ decay modes.
The relative branching ratios of 
$\Gamma_{2\pi^-\pi^+2\pi^0} / \Gamma_{5\pi}$,
$\Gamma_{3\pi^-2\pi^+} / \Gamma_{5\pi}$ and 
$\Gamma_{\pi^-4\pi^0} / \Gamma_{5\pi}$ are plotted against
each other in the three figures.
Note that $\Gamma_{5\pi}$ is set to unity for this figure.
The shaded areas are the regions predicted by the isospin model.
The data points are the experimental results.
The labels indicate the states when 
the width of the state is equal to the total $5\pi$ width
and the width of the other states is zero.}
\label{figure:fivepi}
\end{figure}

The isospin model prediction for 5 pion decay modes is shown in 
fig.~\ref{figure:fivepi}.
The square markers correspond to the solution where the tau decays
exclusively to that state.
The points in fig.~\ref{figure:fivepi} are the experiment results
given in table~\ref{table:br} \cite{pdg98}
and are found to be consistent
with the prediction of the isospin model.
The  data suggests that the tau prefers to decay to the (221) or
$(\omega\rho)$ state, implying 
that the $\taufiveb$ branching ratio should be dominated by the 
$\omga$ decay mode where $\omega \rightarrow \pi^+\pi^-\pi^0$.
The branching ratio of the $\omga$ decay mode is 
$(4.3 \pm 0.5) \times 10^{-3}$ \cite{aleph:br} so that the fraction of 
the $\taufiveb$ branching ratio ($(4.9 \pm 0.4) \times 10^{-3}$
\cite{pdg98}) due to the 
$\omga$ decay is approximately 80\%.
This result is in good agreement with the isospin model prediction.

\begin{figure}
\mbox{\epsfig{file=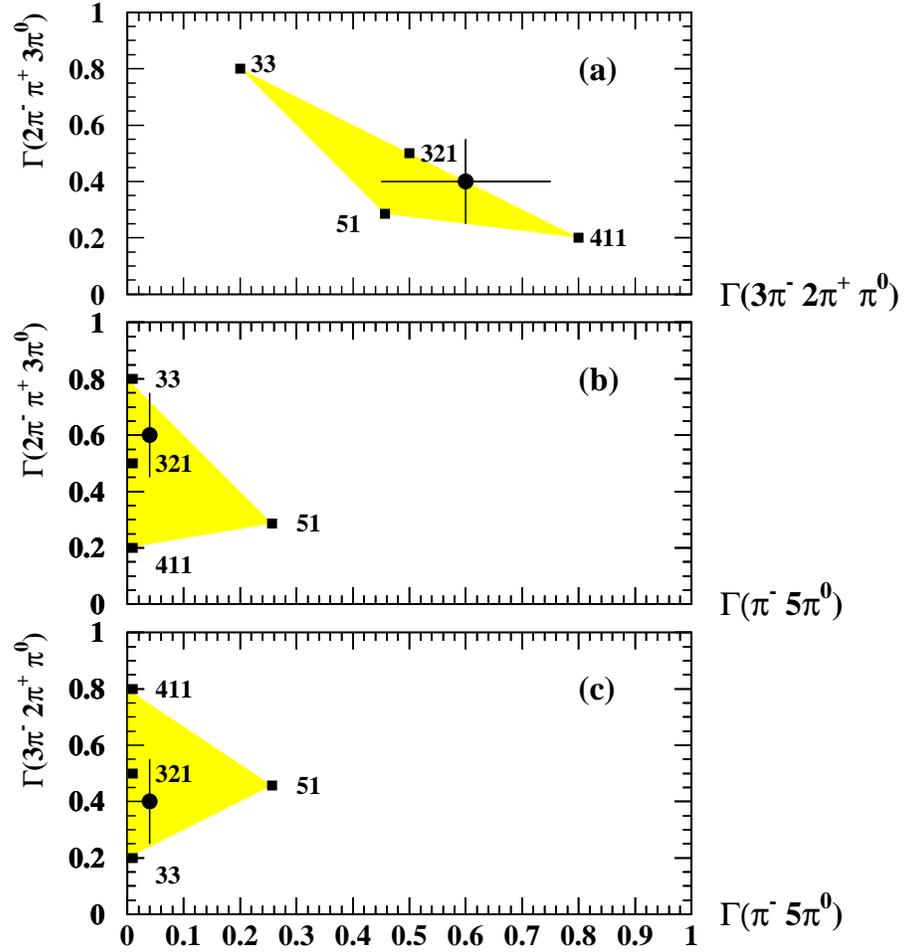,height=15cm}}
\caption[]{The isospin model prediction is plotted for the 6 pion
$\tau$ decay modes.
The relative branching ratios of
$\Gamma_{\pi^-5\pi^0} / \Gamma_{6\pi}$ ,
$\Gamma_{2\pi^-\pi^+3\pi^0} / \Gamma_{6\pi}$ and
$\Gamma_{3\pi^-2\pi^+\pi^0} / \Gamma_{6\pi}$  are plotted against
each other in the three figures.
Note that  $\Gamma_{6\pi}$ is set to unity for this figure.
The shaded areas are the regions predicted by the isospin model.
The data point is the experimental results.
The labels indicate the states when 
the width of the state is equal to the total $6\pi$ width
and the width of the other states is zero.
The data points shown in (b) and (c) have $\Gamma_{\pi^-5\pi^0}=0$
but the points have been offset from the axis for display purposes.
}
\label{figure:sixpi}
\end{figure}

The prediction for the 6 pion tau decays is
given in fig.~\ref{figure:sixpi}.
The experimental values of the 
branching ratios for the 6 pion modes are given in table~\ref{table:br}.
Note that there are no measurements of the $\sixc$ mode.
In fig.~\ref{figure:sixpi}  the relative branching ratios are plotted
assuming that the $\sixc$ branching ratio is zero.
With this assumption, the isospin model suggests that the
(321) or $(\omega\rho\pi$) state should dominate.
In this situation, a large fraction of the $\sixa$ and $\sixb$
branching ratio would be due to the $\omgc$ and $\omgb$ decays, respectively.
If this assumption about the $\sixc$ branching ratio is not correct,
then the fraction from the $\omgc$ and $\omgb$ decays 
would decrease accordingly.

The CLEO Collaboration has measured the $\omgb$ branching ratio
to be  $(1.9 \pm 0.8)\times 10^{-4}$ \cite{cleo:6pion}. 
If the $\omgb$ decay was the only source of the $\sixb$ decay, then
the $\sixb$ branching ratio would be $(1.7 \pm 0.7)\times 10^{-4}$.
Given that the measured value of the $\sixb$ branching ratio is
$(1.44 \pm 0.81)\times 10^{-4}$, one could conclude that the $\sixb$ 
branching ratio is entirely due to the $\omgb$ decay
although the uncertainties on these measurements are quite large.
The CLEO Collaboration in \cite{cleo:6pion} first pointed out 
the importance of the $\omgb$ decay in 6 pion tau decays.
If the $\sixb$ branching ratio is exclusively due to the 
$\omgb$ decay, then the isospin model predicts that the 
$\sixa$ branching ratio would also be entirely due to the $\omgc$ decay.
Unfortunately there are no measurements of the $\omgc$ branching
ratio, however, it should be possible to measure this branching
ratio with existing data sets.

A number of experiments have looked for tau decays to 7 pions
in the $\sevena$  \cite{hrs:7pion, opal:7pion, cleo:7pion}
and  $\sevenb$ modes \cite{cleo:5pion}.
There are six states predicted by the isospin model 
(see table~\ref{table:modes}) and the branching ratios of these states to the
observable final states are given in table~\ref{table:seven}.
If the tau decays to states that include an $\omega$ then 
the (421) or $(\omega\rho 2\pi)$ and the (322) or $(2\omega\pi)$ states
should dominate the 7 pion decays.
This would mean that the $\sevenb$  and $\sevenc$
modes would be the largest 7 pion tau decays and the 
$\sevena$ and $\sevend$ modes would be very much smaller.
As mentioned, there are no experimental measurements of any 7 pion 
branching ratios, however there are limits on the $\sevena$ mode of 
$2.4\times 10^{-6}$ \cite{cleo:7pion} and on the $\sevenb$ mode of 
$1.1\times 10^{-4}$ \cite{cleo:5pion}.
The limit on the $\sevena$ branching ratio is quite small and
hints that the $\omega$ resonance still dominates in the 7 pion channel.

In summary, the isospin model gives a good description of the
5 and 6 pion tau decays.
It is able to predict the large contribution of the $\omega$
resonance in these tau decays.
If the 7 pion tau decays also favour decays via the $\omega$,
then the isospin model could explain the low branching ratio limit
on the $\sevena$ branching ratio.

\noindent
{\it Acknowledgements:} 
I would like to thank Steven Robertson and Richard Keeler for their 
comments and careful reading of the manuscript.
I would also like to acknowledge the support of the 
Institute of Particle Physics of Canada and the
Natural Sciences and Engineering Research Council of Canada.

\begin{table}[ht]
\caption{Isospin states and decay modes for multi-pion tau decays.}
\label{table:modes}
\begin{center}
\begin{tabular}{cc@{\hspace{1cm}}cc@{\hspace{1cm}}cc} \hline
\multicolumn{2}{l}{5 pions} &
\multicolumn{2}{l}{6 pions} &
\multicolumn{2}{l}{7 pions}  \\
$I$-state & State       & $I$-State & State       & $I$-State & State \\ \hline
(500) & $5\pi$          & (510) & $\rho 4\pi$     & (700) & $7\pi$          \\
(410) & $\rho 3\pi$     & (411) & $\omega 3\pi$   & (610) & $\rho 5\pi$     \\
(320) & $2\rho \pi$     & (330) & $3\rho$         & (520) & $2\rho 3\pi$    \\
(221) & $\omega \rho$   & (321) & $\omega\rho\pi$ & (430) & $3\rho \pi$     \\
& & &                                             & (421) & $\omega \rho 2\pi$    \\
& & &                                             & (322) & $2\omega \pi$   \\  \hline
\end{tabular}
\end{center}
\end{table}

\begin{table}
\caption{The branching ratios of the 5 and 6 pion tau decays.
The branching ratios are the fit-value from the Particle Data
Group Compilation \cite{pdg98} and 
exclude the contributions from tau decays involving an
intermediate $K^0$ or $\eta$ resonance.
The third column gives the relative branching ratio for the decays.}
\label{table:br}
\begin{center}
\begin{tabular}{lcc} \hline
Mode         & BR                 & Relative BR \\ \hline
5 pion decays & & \\
$\taufivea$  & $(7.3 \pm 0.8) \times 10^{-4}$ & $0.108 \pm 0.016$ \\
$\taufiveb$  & $(4.9 \pm 0.4) \times 10^{-3}$ & $0.728 \pm 0.067$ \\
$\taufivec$  & $(1.1 \pm 0.6) \times 10^{-3}$ & $0.163 \pm 0.075$  \\ \hline
6 pion decays & & \\
$\sixa$      & $(2.2 \pm 0.5)\times 10^{-4}$   & $0.60 \pm 0.15$ \\
$\sixb$      & $(1.44 \pm 0.81)\times 10^{-4}$ & $0.40 \pm 0.15$ \\
$\sixc$      & no measurement                  & 0 \\ \hline
\end{tabular}
\end{center}
\end{table}

%
\begin{table}
\caption{Isospin coefficients for 7 pion states.}
\label{table:seven}
\begin{center}
\begin{tabular}{lcccc} \hline \noalign{\smallskip}
State    & $\sone$ & $\sthree$ & $\sfive$ & $\sseven$\\ \hline
& & & &   \vspace{-9pt} \\
$(700)$  &  
$\frac{1}{21}$ &  $\frac{4}{35}$  & $\frac{8}{35}$ & $\frac{64}{105}$ \\[0.8ex] 
& & & &   \vspace{-9pt} \\
$(610)$  &  
$\frac{3}{14}$ &  $\frac{3}{14}$  & $\frac{8}{35}$ & $\frac{12}{35}$  \\[0.8ex] 
& & & &   \vspace{-9pt} \\
$(520)$  &  
0              &  $\frac{1}{7}$   & $\frac{18}{35}$ & $\frac{12}{35}$  \\[0.8ex]
& & & &   \vspace{-9pt} \\
$(430)$  &  
0              &  $\frac{2}{5}$   & $\frac{3}{10}$  & $\frac{3}{10}$   \\[0.8ex] 
& & & &   \vspace{-9pt} \\
$(421)$  &  
0              &  $\frac{2}{5}$   & $\frac{3}{5}$   & 0   \\[0.8ex] 
& & & &   \vspace{-9pt} \\
$(322)$  &  
0              &  0               & 1               & 0   \\[0.8ex] 
\noalign{\smallskip}\hline\noalign{\smallskip}
\end{tabular}
\end{center}
\end{table}

\end{document}